\documentclass[aps,twocolumn,showpacs]{revtex4}
\usepackage{bm}
\usepackage{graphicx}
\usepackage{epsfig}
\begin{document}

\title{Optical orientation of electron spins by linearly polarized light}
\author{S.\,A.\,Tarasenko}
\affiliation{A.F.~Ioffe Physico-Technical Institute, Russian
Academy of Sciences, 194021 St.~Petersburg, Russia }
\begin{abstract}
Absorption of circularly polarized light in semiconductors is
known to result in optical orientation of electron and hole spins.
It has been shown here that in semiconductor quantum well
structures spin orientation of carriers can be achieved by
linearly or even unpolarized light. Moreover, the sign and
magnitude of the spin orientation can be varied by rotating the
polarization plane of incidence light. The effect under study is
related to reduced symmetry of the quantum wells as compared to
bulk materials and, microscopically, caused by zero-field spin
splitting of electron and hole states.
\end{abstract}

% 72.25.-b Spin polarized transport
% 72.25.Fe Optical creation of spin polarized carriers
% 72.25.Rb Spin relaxation and scattering
% 72.25.Dc Spin polarized transport in semiconductors
% 78.67.De Optical properties of Quantum wells

\maketitle

Spin-dependent phenomena in semiconductor structures are the
subject of extensive ongoing research. One of the most widespread
and powerful methods for creating spin polarization and
investigating kinetics of spin-polarized carriers is optical
orientation of electron and nuclear spins by circularly polarized
light~\cite{oo,IT_JETP,Pfalz,Sanada,Braun}. This effect can be
interpreted as a transfer of the photon angular momenta to free
carriers. Under interband excitation by circularly polarized
light, direct optical transitions from the valence band to the
conduction band can occur only if the electron angular momentum is
changed by $\pm1$. These selection rules lead to the spin
orientation of photoexcited carriers, with degree and sign of spin
orientation depending on the light helicity.

In the present paper we show that in low-dimensional semiconductor
systems spin orientation of carriers can be achieved by linearly
or even unpolarized light. The effect under consideration is
related to reduced symmetry of the low-dimensional structures as
compared to bulk compounds and is forbidden in bulk cubic
semiconductors. Microscopically, it is caused by asymmetrical
photoexcitation of carriers in spin subbands followed by spin
precession in an effective magnetic field induced by the Rashba or
Dresselhaus spin-orbit coupling~\cite{Ganichev}.

The effect is most easily conceivable for direct transitions
between the heavy-hole valence subband $hh1$ and the conduction
subband $e1$ in quantum well (QW) structures of the C$_s$ point
symmetry, e.g. in (113)- or (110)-grown QWs based on
zinc-blende-lattice compounds. In such structures the spin
component along the QW normal $z$ is coupled with the in-plane
electron wave vector. This leads to $\bm{k}$-linear spin-orbit
splitting of the energy spectrum as sketched in Fig.~1, where the
heavy-hole subband $hh1$ is split into two spin branches $\pm 3/2$
shifted relative to each other in the $\bm{k}$ space. Due to the
selection rules the allowed optical transitions from the valence
subband $hh1$ to the conduction subband $e1$ are $|+3/2 \rangle
\rightarrow |+1/2 \rangle$ and $|-3/2 \rangle \rightarrow |-1/2
\rangle$, as illustrated in Fig.~1 by dashed vertical lines.
\begin{figure}[t]
\leavevmode \epsfxsize=0.85\linewidth
\centering{\epsfbox{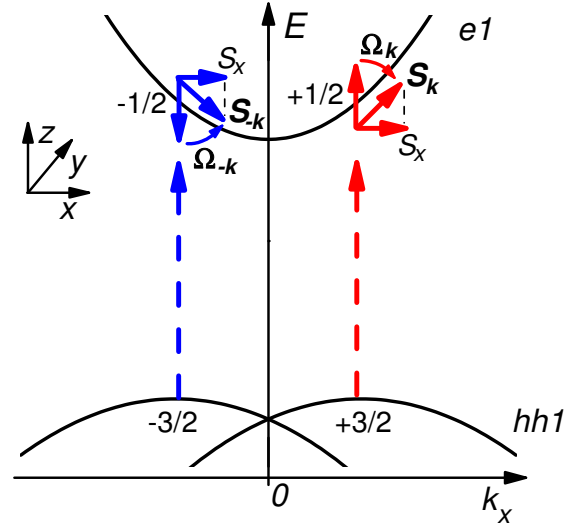}} \caption{(Color online).
Microscopic origin of the optical orientation of electrons spins
by linearly polarized light. Asymmetry of photoexcitation followed
by spin precession leads to appearance of average electron spin.
The vertical dashed lines show the possible optical transitions.
Spin-orbit coupling in the conduction subband is taken into
account here as an effective magnetic field acting on electron
spins.} \vspace{-0.5cm}
\end{figure}
Under excitation with linearly polarized or unpolarized light the
rates of both transitions coincide. In the presence of the spin
splitting, the optical transitions induced by photons of the fixed
energy $\hbar\omega$ occur in the opposite points of the $\bm{k}$
space for the electron spin states $\pm 1/2$. Such an asymmetry of
photoexcitation results in non-equilibrium distribution where
electrons with spin $+1/2$ propagate mainly in one direction, e.g.
$k_x >0$, and those with the spin $-1/2$ propagate in the opposite
direction, $k_x <0$~\cite{TI_JETPLett,Sherman}. Spin-orbit
interaction is known to be present and couple spin states and
in-plane movement of free carriers in conduction subbands as well.
The spin-orbit coupling can be considered as an effective magnetic
field $\bm{B}_{\bm k}$ acting on electron spins, with the field
direction depending on the electron wave vector $\bm k$ and its
strength being proportional to $|\bm k|$~\cite{DP,DK,AGW}.
Spin-dependent asymmetry of photoexcitation considered above is
caused by spin-orbit interaction in both the valence and
conduction subbands and, in general, does not correspond to eigen
state of the spin-orbit coupling in the subband $e1$. Therefore,
electron spins originally directed, according to the selection
rules, along or opposite to the QW normal will precess in the
effective magnetic field $\bm{B}_{\bm k}$~\cite{Kalevich}, which
has non-zero in-plane component, as shown in Fig.~1. Electrons
with the initial spin $+1/2$ and wave vector $k_x>0$ are affected
by effective field with the Larmor frequency $\bm{\Omega}_{\bm
k}$, while carriers with the opposite spin, $-1/2$, and opposite
wave vector, $-k_x$, are affected by field with the frequency
$\bm{\Omega}_{-\bm k}$. Since in QW structures the effective
magnetic field caused by spin-orbit coupling is linear in the wave
vector, then $\bm{\Omega}_{-\bm k}=-\bm{\Omega}_{\bm k}$ and the
rotation axes are opposite for carriers with the initial spins
$\pm 1/2$. As a result, the precession leads to an appearance of
spin component $S_x>0$ for carriers with both positive and
negative $k_x$ as shown in Fig.~1. The value of the generated
electron spin is determined by the average angle of spin rotation.
Thus, interband absorption of unpolarized light in QW structures
of low-symmetry results in spin orientation of photoinduced
carriers. The spin polarization of electron gas disappears after
photoexcitation with the conventional spin relaxation time.

Generally, direction of the optically oriented electron spins is
determined by light polarization and explicit form of spin-orbit
interaction in both the conduction and valence bands. The latter
is governed by the QW symmetry and can be varied. In QWs based on
zinc-blende-type semiconductors and grown along the
crystallographic direction $z \| [110]$, absorption of unpolarized
light leads to orientation of electron spins in the QW plane,
along $x \| [1\bar{1}0]$. We assume the relaxation time of the
asymmetrical electron distribution $\tau_e$ to be shorter than the
Larmor precession period, $\Omega_{\bm{k}} \tau_e \ll 1$. Then,
generation rate of the spin component determined by the average
angle of electron spin rotation can be estimated as
\begin{equation}
\dot{S}_x = \frac{1}{2} \, \tau_e \Omega_{y, \bm{k}_0} \dot{N} \:,
\end{equation}
where $\Omega_y$ is the $y$-component of the Larmor frequency of
the effective magnetic field, $y \| [001]$, $\bm{k}_{0}$ is the
average wave vector of electrons with the spin projection $+1/2$
along the QW normal in the moment of photoexcitation, and
$\dot{N}$ is generation rate of electrons in the subband $e1$.
Taking into account the explicit form of the spin-orbit
interaction in (110)-grown QWs of the $C_s$ point-group symmetry,
one derives
\begin{equation}
\dot{S}_x = \gamma_{yx}^{(e1)} \gamma_{zx}^{(hh1)}
\frac{\mu_{e,hh} \tau_e}{\hbar^3} \dot{N} \:.
\end{equation}
Here $\gamma_{\alpha\beta}^{(\nu)}$ ($\nu=e1,hh1$) are the
constants describing linear in the wave vector coupling between
$\alpha$-component of the electron angular momentum and
$\beta$-component of the wave vector in the subbands $e1$ and
$hh1$, respectively, $\alpha$ and $\beta$ are the Cartesian
coordinates, $\mu_{e,hh} = m_e m_{hh}^{\|} / (m_e + m_{hh}^{\|})$
is the reduced mass, $m_e$ and $m_{hh}^{\|}$ are the electron and
heavy hole effective masses in the QW plane, respectively.

Possibility to achieve optical orientation by linearly polarized
light in various low-dimensional structures follows also from
symmetry analysis. Phenomenologically, spin generation by light is
described by
\begin{equation}
\dot{S}_{\alpha}= I \sum_{\beta\gamma} \chi_{\alpha\beta\gamma}\,
\frac{e_{\beta} e_{\gamma}^* + e_{\gamma} e_{\beta}^*}{2}
 + I \sum_{\beta} \phi_{\alpha\beta} \, \mathrm{i} [\bm{e} \times
\bm{e}^*]_{\beta} \:,
\end{equation}
where $\dot{S}_{\alpha}$ are the generation rates of the spin
components, $I$ is the light intensity, $\bm{e}$ is the (complex)
unit vector of the light polarization, $\bm{e}^*$ is the vector
complex conjugated to $\bm{e}$. The pseudo-tensor
$\phi_{\alpha\beta}$ describes ``conventional'' optical
orientation by circularly polarized light since the vector product
$\mathrm{i}[\bm{e} \times \bm{e}^*]$ is proportional to the light
helicity and vanishes for linearly polarized light. In contrast,
the symmetrized product $(e_{\beta} e_{\gamma}^* + e_{\gamma}
e_{\beta}^*)/2$ is insensitive to the light helicity for
elliptically polarized radiation and reaches maximum for linear
polarization. Thus, the third-rank tensor
$\chi_{\alpha\beta\gamma}$, symmetrical in the last two indices,
$\chi_{\alpha\beta\gamma}=\chi_{\alpha\gamma\beta}$, describes
spin orientation by linearly polarized light. In what follows we
consider this effect and assume the polarization vector $\bm e$ to
be real.

Symmetry analysis shows that in zinc-blende- or diamond-type bulk
crystals, $T_d$ and $O_h$ point groups, respectively, all
components of $\chi_{\alpha\beta\gamma}$ vanish, and optical
orientation of electron and hole spins can be achieved by
circularly polarized light only. In contrast, in low-dimensional
systems grown on the basis of cubic semiconductors, non-zero
components of $\chi_{\alpha\beta\gamma}$ do exist, allowing for
spin orientation by light of zero helicity. In particular, in QWs
of the C$_s$ symmetry the tensor $\chi_{\alpha\beta\gamma}$
contains 8 independent constants, and spin orientation can be
achieved even under excitation with unpolarized light as was
demonstrated above.

In (001)-grown QWs spin orientation can not be achieved by
unpolarized light, but is allowed under excitation with linearly
polarized light. Asymmetrical (001)-grown structures, such as
single heterojunctions or QWs with non-equivalent normal and
inverted interfaces, belong to the $C_{2v}$ point-group symmetry,
and optical orientation by linearly polarized light is described
here by three independent constants $A$, $B$ and $C$ as follows
\begin{equation}
\dot{S}_{z'} = A e_{x'} e_{y'} \;,\;\; \dot{S}_{x'} = B e_{y'}
e_{z'} \;,\;\; \dot{S}_{y'} = C e_{x'} e_{z'} \;,
\end{equation}
where $z' \| [001]$ is the QW normal, $x' \| [1\bar{1}0]$ and
$y'\|[110]$. One can see that excitation with linearly polarized
light under normal incidence may result in orientation of electron
spins along the QW normal, with the sign and magnitude depending
on the light polarization. The point-group symmetry of (001)-grown
QWs with equivalent interfaces is enhanced to $D_{2d}$ which
allows only one linearly independent constant: $A=0$, $B=-C$.
Particularly, it follows that in such structures excitation with
linearly polarized light in the geometry of normal incidence does
not lead to spin orientation. In the other limiting case, when the
spin-orbit coupling is determined only by the structure inversion
asymmetry unrelated to the crystal lattice, as it can happen in
QWs grown of centrosymmetrical semiconductor compounds like SiGe,
the symmetry of the structure is effectively increased to
$C_{\infty v}$ and the relations $A=0$, $B=-C$ retain. Thus,
generation of electron spins along the QW normal is possible in
asymmetrical (001)-grown QWs, but vanishes for symmetrical
structures of the $D_{2d}$ point group as well as for uniaxial
structures of the $C_{\infty v}$ symmetry.

A consistent theory of the spin orientation by linearly polarized
light is conveniently developed by using the spin density matrix
technique. Dynamics of the density matrix $\rho$ of photoexcited
electrons  in the subband $e1$  is described by equation~\cite{oo}
\begin{equation}\label{rho}
\frac{\partial \rho}{\partial t} + \frac{\rho}{\tau_0} +
\frac{i}{\hbar}[H_{so}^{(e1)},\rho] = G + \mathrm{St} \rho \:.
\end{equation}
Here $\tau_0$ is the lifetime of photoelectrons, $H_{so}^{(e1)}$
is the spin-orbit contribution to the Hamiltonian,
\begin{equation}
H_{so}^{(e1)} =  \frac{\hbar}{2} (\bm{\Omega}_{\bm{k}}^{(e1)}
\cdot \bm{\sigma}) \:,
\end{equation}
$\bm{\Omega}_{\bm{k}}^{(e1)}$ is the Larmor frequency of the
spin-orbit coupling-induced effective magnetic field,
$\bm{\sigma}$ is the vector composed of the Pauli matrices
$\sigma_x$, $\sigma_y$ and $\sigma_z$, $G$ is the matrix of
electron photogeneration, and $\mathrm{St} \rho$ is the collision
integral that describes electron scattering by phonons, static
defects, charge carriers, etc., leading to equilibration. It is
convenient to expand the density matrix $\rho$ and the matrix of
photogeneration $G$ into diagonal and spin components as follows
\[
\rho = f_0 I + (\bm{S}_{\bm{k}} \cdot \bm{\sigma}) \:,
\]
\vspace{-0.7cm}
\[
G = g_0 I + (\bm{g}_{\bm{k}} \cdot \bm{\sigma}) \:, \nonumber
\]
where $f_0=\mathrm{Tr}\rho/2$ is the distribution function of
electrons, $\bm{S}_{\bm{k}} = \mathrm{Tr} (\bm{\sigma} \rho)/2$ is
the total spin of electrons with the wave vector $\bm k$, $2g_0$
is the rate of carrier photogeneration, $\bm{g}_{\bm{k}}$ is the
rate of spin photogeneration into the state with the wave vector
$\bm k$, and $I$ is the $2 \times 2$ unit matrix. Then, for
steady-state regime, the equation for the spin density
$\bm{S}_{\bm k}$ in the relaxation time approximation and
neglecting spin-flip scattering has the form
\begin{equation}\label{S_k}
\frac{\bm{S}_{\bm{k}}}{\tau_0} + [\bm{S}_{\bm{k}} \times
\bm{\Omega}_{\bm{k}}^{(e1)}] = \bm{g}_{\bm{k}} -
\frac{\bm{S}_{\bm{k}} - \bar{\bm{S}}_{\bm{k}} }{\tau_e} \:,
\end{equation}
where $\bar{\bm{S}}_{\bm{k}}$ is $\bm{S}_{\bm{k}}$ averaged over
directions of the wave vector $\bm k$, $\tau_e$ is the
isotropisation time of the spin density $\bm{S}_{\bm{k}}$. In the
case of elastic scattering by static defects in two-dimensional
structures, the time $\tau_e$ coincides with the conventional
momentum relaxation time that governs the electron mobility.
However, we note that electron-electron collisions between
particles of opposite spins, which do not affect the mobility, can
contribute to relaxation of the asymmetrical spin-dependent
distribution and decrease the time $\tau_e$, as it happens, e.g.,
in spin relaxation~\cite{Glazov}. Assuming the value $\tau_e
\Omega_{\bm k}^{(e1)}$ to be a small parameter, the solution of
Eq.~(\ref{S_k}) for the spin density $\bar{\bm{S}}_{\bm{k}}$ to
the second order in $\Omega_{\bm k}^{(e1)}$ has the form
\begin{equation}\label{S_average}
\frac{\bar{\bm{S}}_{\bm{k}}}{\tau_0}  + \tau_e
\overline{[\bm{\Omega}_{\bm{k}}^{(e1)} \times
[\bar{\bm{S}}_{\bm{k}} \times \bm{\Omega}_{\bm{k}}^{(e1)}]]} =
\bar{\bm g}_{\bm k} + \tau_e
\overline{[\bm{\Omega}_{\bm{k}}^{(e1)} \times \bm{g}_{\bm{k}}]}
\:,
\end{equation}
where the overline means averaging over directions of the wave
vector. The first term in the left-hand side of
Eq.~(\ref{S_average}) describes disappearance of the total
electron spin due to recombination. The second term in the
left-hand side is responsible for the D'yakonov-Perel' spin
relaxation mechanism~\cite{DP,DK}. The right-hand side of
Eq.~(\ref{S_average}) describes orientation of electron spins. The
first term is responsible for ``conventional'' optical orientation
by circularly polarized light, while the second term describes
spin generation caused by asymmetric photoexcitation $\bm{g}_{\bm
k}$ followed by spin precession in effective magnetic field with
the Larmor frequency $\bm{\Omega}_{\bm{k}}$. Under illumination
with linearly polarized or unpolarized light $\bar{\bm g}_{\bm k}$
is zero, and the spin generation is given by the second term.
Then, the total spin generation rate in the subband $e1$ has the
form
\begin{equation}\label{spin_gen}
\dot{\bm{S}} = \sum_{\bm{k}} \tau_e [\bm{\Omega}_{\bm{k}}^{(e1)}
\times \bm{g}_{\bm{k}}] \:.
\end{equation}

As an example, let us consider optical orientation of electron
spins in (001)-grown QWs under normal incidence of linearly
polarized light. In contrast to the energy spectrum in
low-symmetry structures sketched in Fig.~1, in (001)-QWs the
$\bm{k}$-linear spin splitting of the $hh1$ valence subband is
depressed and here, for the sake of simplicity, we consider
optical transitions between the light-hole subband $lh1$ and the
conduction subband $e1$. Calculations show that in this particular
case the dependence of the photogeneration matrix components on
the polarization vector $\bm e$ to the first order in the
spin-orbit interaction has the form
\begin{eqnarray}\label{generation}
\bm{g}_{\bm{k}} &=& \left[ \bm{\Omega}_{\bm{k}}^{(e1)} + 2 \bm{e}
(\bm{\Omega}_{\bm{k}}^{(lh1)} \cdot \bm{e}) -
\bm{\Omega}_{\bm{k}}^{(lh1)} \right] \frac{\hbar}{2}
\frac{\partial g_0}{\partial \varepsilon_k} \:, \\
g_0 &=& \frac{\pi}{3\hbar} \left(\frac{eA}{c m_0}\right)^2
|P_{cv}|^2 \, \delta(E_{e1,\,lh1} + \varepsilon_k - \hbar\omega)
\:.\nonumber
\end{eqnarray}
Here $e$ is the electron charge, $A$ is the amplitude of the
vector potential of the light wave, $c$ is the light velocity,
$m_0$ is the free electron mass, $P_{cv}=\langle S|\hat{p}_z|Z
\rangle$ is the interband matrix element of the momentum operator,
$E_{e1,\,lh1}$ is the energy gap between the subbands $lh1$ and
$e1$, $\varepsilon_k = \hbar^2 k^2 / 2 \mu_{e,lh}$, and
$\mu_{e,lh}=m_e m_{lh}^{\|} / (m_e + m_{lh}^{\|})$ is the reduced
mass for the in-plane motion.

In (001)-grown structures the vectors
$\bm{\Omega}^{(\nu)}_{\bm{k}}$ have the form
\[
\bm{\Omega}_{\bm{k}}^{(\nu)} = \frac{2}{\hbar}
(\gamma_{x'y'}^{(\nu)} k_{y'}, \gamma_{y'x'}^{(\nu)} k_{x'}, 0)
\:,
\]
where $\nu=e1, lh1$ is the subband index. Then, substituting
Eq.~(\ref{generation}) into Eq.~(\ref{spin_gen}), one derives the
spin generation rate in the subband $e1$
\begin{equation}\label{S_z}
\dot{S}_{z'} = 2 e_{x'} e_{y'} \left( \gamma_{y'x'}^{(e1)}
\gamma_{y'x'}^{(lh1)} - \gamma_{x'y'}^{(e1)} \gamma_{x'y'}^{(lh1)}
\right) \frac{\mu_{e,lh} \tau_e}{\hbar^3} \dot{N} \:.
\end{equation}

Optical orientation of electron spins by linearly polarized light
can be observed and studied with conventional technique for
detection of spin orientation, e.g. by analyzing circular
polarization of luminescence under electron-hole radiative
recombination. Moreover, the dependence of $\dot{S}_{z'}$ on
polarization of the incident light given by Eq.~(\ref{S_z}) allows
one to separate the effect under study from possible experimental
background noise. Indeed, the spin generation $\dot{S}_{z'}$ is of
opposite sign for the exciting light polarized along the $[100]$
and $[010]$ crystallographic axes and vanishes for the light
polarized along the $[1\bar{1}0]$ or $[110]$ axes. Generally, the
dependence of the spin orientation on the light polarization is
given by $\dot{S}_{z'} \propto 2 e_{x'} e_{y'} = \sin 2\varphi$,
where $\varphi$ is the angle between the light polarization plane
and the $[1\bar{1}0]$ axis.

The spin generation rate given by Eq.~(\ref{S_z}) is proportional
to constants of the spin-orbit coupling in both $e1$ and $lh1$
subbands and vanishes if the product $\gamma_{y'x'}^{(e1)}
\gamma_{y'x'}^{(lh1)}$ equals to $\gamma_{x'y'}^{(e1)}
\gamma_{x'y'}^{(lh1)}$. Appearance of the $\bm k$-linear terms is
connected with reduction of the system symmetry as compared to
bulk materials. In (001)-grown QWs based on zinc-blende-lattice
semiconductors, there are two types of the $\bm k$-linear
contributions to the effective Hamiltonians of electron and
light-hole subbands~\cite{Ganichev,AGW}. The contributions can
originate from the lack of an inversion center in the bulk
compositional semiconductors or/and from anisotropy of chemical
bonds at the interfaces (so-called Dresselhaus term)~\cite{DK},
and can be induced by the heterostructure asymmetry unrelated to
the crystal lattice (Rashba term)~\cite{Rashba2}. The constants
describing $\bm k$-linear spin splitting in the subbands $e1$ and
$lh1$ are related to the corresponding Dresselhaus and Rashba
constants by
\begin{eqnarray}
\gamma_{x'y'}^{(\nu)} = \gamma_D^{(\nu)} + \gamma_R^{(\nu)} \:,\\
\gamma_{y'x'}^{(\nu)} = \gamma_D^{(\nu)} - \gamma_R^{(\nu)} \:.
\nonumber
\end{eqnarray}
In symmetrical (001)-grown QWs, the spin-orbit coupling is given
by the Dresselhaus term only, while the Rashba term vanishes. In
this case the constants $\gamma_{x'y'}^{(\nu)}$ and
$\gamma_{y'x'}^{(\nu)}$ are equal and, hence, the expression in
the parenthesis in Eq.~(\ref{S_z}) is zero and the spin
orientation does not occur. In the other limiting case, if the
Rashba coupling dominates and the Dresselhaus term is negligible,
the constants are related by $\gamma_{x'y'}^{(\nu)} =
-\gamma_{y'x'}^{(\nu)}$, and the effect vanishes as well. Thus,
only in QWs with both the Rashba and the Dresselhaus spin-orbit
couplings one can induce orientation of electron spins by normally
incident linearly polarized light. This result is in full
agreement with the symmetry analysis presented above. The spin
orientation along the QW normal by linearly polarized light is
possible for asymmetrical (001)-grown QWs, but vanishes for
symmetrical structures of the $D_{2d}$ class as well as for
uniaxial structures of the $C_{\infty v}$ symmetry.

The constants of spin-orbit coupling and the relaxation time can
be estimated as $\gamma/\hbar \sim 10^{5}$~cm/s, $\tau_e \sim
10^{-11}$~s. Then, an estimation for electron spin generated under
absorption of one photon following Eq.~(\ref{S_z}) gives $\dot{S}
/ \dot{N} \sim 10^{-2}$ (or $1\%$).

Spin orientation of carriers, caused by asymmetrical
photoexcitation followed by spin precession in the effective
magnetic field, can be achieved not only under interband optical
transitions, but also under intersubband and intrasubband
(Drude-like) photoexcitation in QW structures. In the latter case
it can be considered as a nonlinear effect of generation of
\textit{dc} spin polarization by \textit{ac} electric field.

It should be noted that circular polarization of luminescence
under excitation with linearly polarized light in zero magnetic
field was observed under study of excitons localized on
anisotropic islands in QWs~\cite{exciton}. This effect is caused
by optical alignment of exciton dipoles followed by dipole
oscillations in anisotropic media and, generally speaking, can be
observed in spinless systems.

The author acknowledges helpful discussions with E.L.~Ivchenko.
This work was supported by the RFBR, programs of the RAS, and
Foundation ``Dynasty'' - ICFPM.

\end{document}